\documentclass[12pt]{article}

\pdfoutput=1

\usepackage{graphics}
\usepackage{amssymb}
\usepackage{amsmath}
\usepackage{tikz}

\newcommand{\beq}{\begin{equation}}

\newcommand{\eeq}{\end{equation}}
\newcommand{\bea}{\begin{eqnarray}}
\newcommand{\eea}{\end{eqnarray}}

\begin{document}

\begin{center}
${}$\\
\vspace{100pt}
{ \Large \bf Quantizing Quantum Ricci Curvature 
}

\vspace{46pt}

{\sl N. Klitgaard}
and {\sl R. Loll}

\vspace{24pt}
{\footnotesize

Institute for Mathematics, Astrophysics and Particle Physics, Radboud University \\ 
Heyendaalseweg 135, 6525 AJ Nijmegen, The Netherlands.\\ 
{email: n.klitgaard@science.ru.nl, r.loll@science.ru.nl}\\
}
\vspace{48pt}

\end{center}

\vspace{0.8cm}

\begin{center}
{\bf Abstract}
\end{center}

\noindent 
Quantum Ricci curvature has been introduced recently as a new, geometric observable characterizing the curvature
properties of metric spaces, without the need for a smooth structure. Besides coordinate invariance, its key 
features are scalability, computability and robustness. We demonstrate that these properties continue to hold
in the context of nonperturbative quantum gravity, by evaluating the quantum Ricci curvature numerically
in two-dimensional Euclidean quantum gravity, defined in terms of
dynamical triangulations. Despite the well-known, highly nonclassical properties of the underlying quantum geometry,
its Ricci curvature can be matched well to that of a five-dimensional round sphere.

\vspace{12pt}
\noindent


\newpage

\section{Quantum gravity needs more observables}
\label{sec:need}

Suppose quantum gravity and spacetime at the Planck scale can be described by 
local dynamical degrees of freedom, which are geometric in the sense that (quantum analogues of)
geodesic distances and volumes exist. We are primarily interested in situations where, unlike in general relativity, there 
is no a priori differentiable manifold $M$ and smooth metric structure $g_{\mu\nu}(x)$ present, and standard tensor calculus is 
therefore not available. Can we nevertheless describe quantum spacetime in terms of generalized geometry, and on sufficiently
large scales recover the usual notions of classical geometry? 

Quantum Gravity from Causal Dynamical Triangulations (CDT) is a framework of this kind, where moreover such questions can be 
addressed explicitly and quantitatively, and some have already been answered in the affirmative (see \cite{physrep} for a review).  
Roughly speaking, ``quantum spacetime" in this setting arises from a suitable scaling limit of a gravitational path integral.
The latter takes the form of a weighted superposition of piecewise 
flat spacetimes, which serve as regulated approximations of curved spacetime geometries of a given fixed topology.  

Any analysis of the physical content of such a candidate theory of quantum gravity has to be made in terms of observables. 
Observables are purely geometric functions of the dynamical variables that do not depend on spacetime labels, a
requirement that in classical gravity is usually phrased as `invariance under diffeomorphisms'. However, the observation
that `a scalar function $\phi(x)$ cannot be an observable because it depends on the spacetime label $x$' is
independent of whether $x$ takes discrete, continuous or smooth values. As is well known, observables
in pure gravity are highly non-local and typically take the form of integrals or averages of scalar quantities 
over spacetime, a feature that persists in the quantum theory.  

The general challenge in nonperturbative quantum gravity without a background metric structure, smooth or otherwise, 
is to construct what we will call {\it good quantum observables}.
Apart from being geometric, they are defined to be scalable, that is, associated with a variable
distance scale $\delta$, and their expectation values should be computable, finite and not identically zero. Lastly, they should
have a well-defined classical limit for sufficiently large $\delta$. Although for small $\delta$ such an observable 
${\cal O}(\delta)$ will characterize spacetime properties at short distances, it will not be local in the usual sense, 
because the definition of ${\cal O}(\delta)$ will typically involve an integration over spacetime.

The quite striking results to come out of CDT quantum gravity,  
including the emergence of de Sitter space \cite{desitter}, the crucial role of causal structure \cite{CDT1} and the phenomenon of 
dynamical dimensional reduction \cite{spectral} 
have been based on the study of just four geometric observables: the spectral and Hausdorff dimensions, 
the total spatial volume as a function of cosmological proper time (i.e. the global shape of the universe),
and quantum fluctuations of the latter. However, more observables are clearly needed, for example,
to understand the nature of the newly found phase transition \cite{bifurcation}, to improve the existing renormalization group 
analysis \cite{cdtrg} and to quantify (Planckian) physics near the already found second-order phase transitions \cite{secondorder}.

To improve on this situation, we recently introduced a new observable, the {\it quantum Ricci 
curvature}, and investigated some of its properties on smooth and piecewise flat classical spaces \cite{qrc1}. Building on
these results, which will be summarized in Sec.\ \ref{sec:qrc}, we present in Sec.\ \ref{sec:dtqg} the first genuine quantum implementation
of quantum Ricci curvature, namely, in two-dimensional quantum gravity formulated in terms of dynamical triangulations (DT),
the Euclidean precursor of CDT (see \cite{bookadj} for an introduction).  

DT quantum gravity is an excellent testing ground for our prescription. Firstly, in the context of nonperturbative
quantum gravity, the important role of the {\it Hausdorff dimension} in characterizing the universal, intrinsic 
properties of quantum geometry was highlighted first in pioneering work in two-dimensional DT \cite{dim2d,dim2dnum}. 
This quantity is a prime example of a good quantum observable in our sense, in any dimension.
Secondly, as also revealed by this early work, the quantum geometry obtained in this model is fractal and
highly nonclassical (with Hausdorff dimension $d_H\! =\! 4$), and its behaviour is dictated by the dynamics of branching baby 
universes on all scales.
It will be particularly interesting to understand to what extent the quantum Ricci curvature will reflect the
pure quantum nature of the underlying geometry.

\section{Quantum Ricci curvature}
\label{sec:qrc}

Since Ricci curvature in the continuum is a two-tensor, the question arises how its
inherent directional dependence can be captured in a non-smooth context. 
We will associate the Ricci curvature $Ric(v,v)=R_{ij}v^iv^j$ along a vector $v$ with a quasi-local
construction involving a pair of overlapping geodesic spheres.
Note that because of its symmetry the Ricci tensor $R_{ij}(x)$ is determined completely by evaluating it on 
pairs of identical vectors. 

\begin{figure}[t]
\centerline{\scalebox{0.45}{\rotatebox{0}{\includegraphics{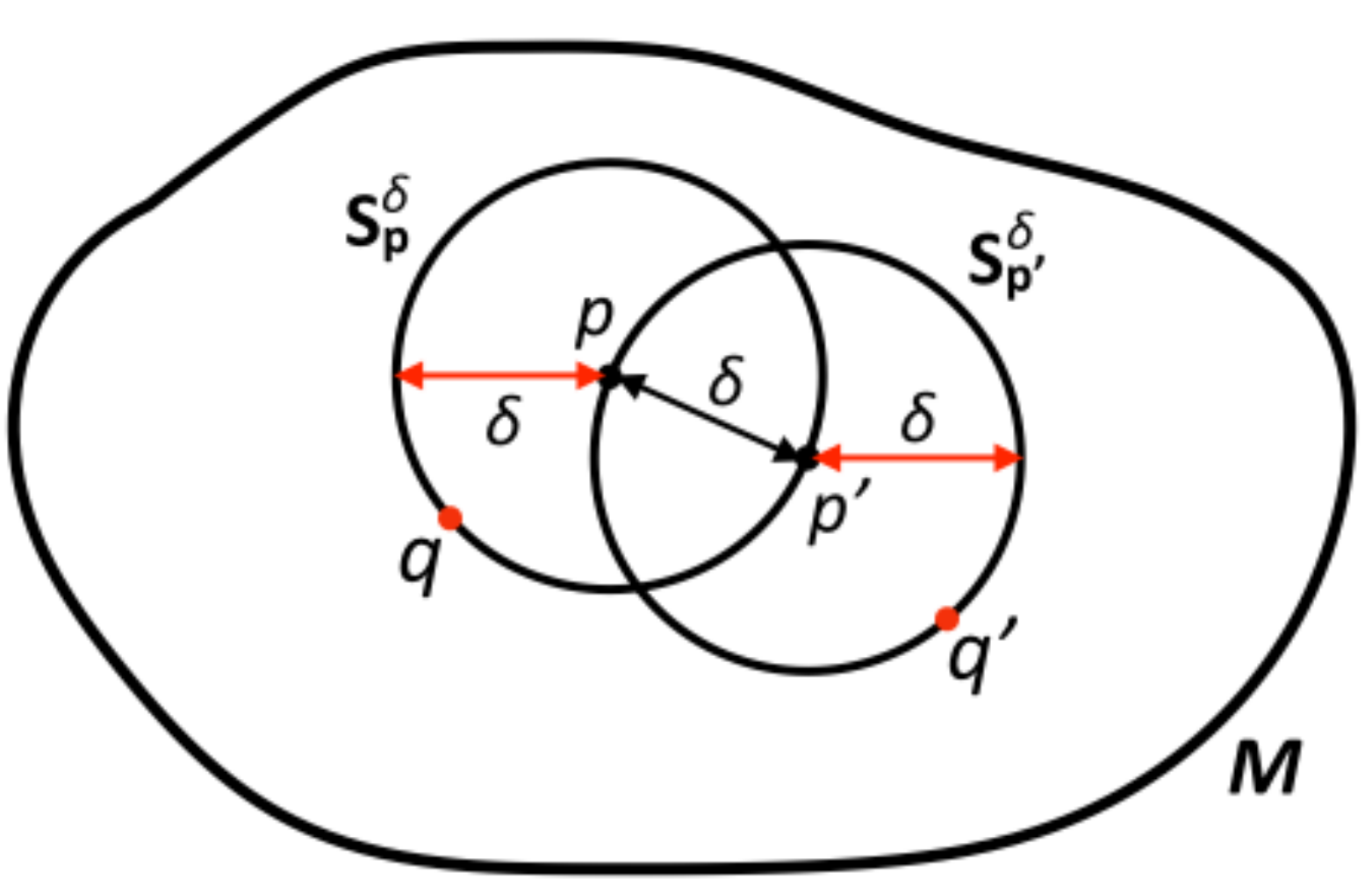}}}}
\caption{Two overlapping $\delta$-spheres, whose centres $p$ and $p'$ are a distance $\delta$ apart. Determing the
quasi-local quantum Ricci curvature associated with this configuration involves averaging over the distances between all point
pairs $(q,q')$ along the circles.}
\label{fig:intersectingspheres}
\end{figure}

Our definition of quantum Ricci curvature \cite{qrc1} is inspired by Ollivier's coarse Ricci curvature \cite{ollivier}, but uses
a different notion of sphere distance, which is essential for the application in quantum gravity.\footnote{Note that our construction is applicable to
metric spaces of positive definite signature, such as what one encounters in CDT quantum gravity after applying the ``Wick rotation"
that exists in this formulation \cite{physrep}.} 
The crucial quantity to compute is the average sphere distance $\bar d$
of two overlapping spheres $S_p^{\delta}$ and $S_{p'}^{\delta}$ of radius $\delta$, whose centres 
$p$ and $p'$ are a geodesic distance $\delta$ apart, as illustrated in Fig.\ \ref{fig:intersectingspheres}.
On a $D$-dimensional Riemannian manifold, for sufficiently small $\delta$, $\bar d$ is given by the normalized integral over
the two spheres,
\begin{equation}
\bar{d}(S_p^{\delta},S_{p'}^{\delta}):=\frac{1}{vol(S_p^{\delta})}\frac{1}{vol(S_{p'}^{\delta})}
\int_{S_p^{\delta}}d^{D-1}q\; \sqrt{h} \int_{S_{p'}^{\delta}}d^{D-1}q'\; \sqrt{h'}\ d(q,q'),
\label{sdist}
\end{equation}   
where $h$ and $h'$ are the determinants of the metrics induced on $S_p^{\delta}$ and $S_{p'}^{\delta}$, 
and $d(q,q')$ denotes the geodesic distance between the points $q$ and $q'$. 
The important feature of eq.\ (\ref{sdist}) is that it can be generalized to non-smooth metric spaces and be computed
in a straightforward way. A $\delta$-sphere $S^\delta_p$ around $p$ will in this case be interpreted as the set of all points $q$ with distance 
$\delta$ from $p$.  
Given $\bar{d}$, the dimensionless quantum Ricci curvature $K_q(p,p')$ is defined via 
\begin{equation}
\frac{\bar{d}(S_p^{\delta},S_{p'}^{\delta})}{\delta}=c_q (1 - K_q(p,p')),\;\;\; \delta =d(p,p'),
\label{qric}
\end{equation}
where $c_q$ is a positive constant, which depends on the metric space under consideration,
and $K_q$ captures any nontrivial dependence on $\delta$. 

Although our interest is in the non-smooth case and non-infinitesimal $\delta$, expression (\ref{qric}) can just as well be evaluated 
on a general smooth $D$-dimensional Riemannian manifold in the limit as $\delta\rightarrow 0$.
Using Riemann normal coordinates based at the point $p$, say, and
expressing the result as a power series in $\delta$, one finds in dimensions $D\!=\! 2$, 3 and 4
\begin{eqnarray}
\frac{\bar d}{\delta}=
\left\{\begin{array}{c}
1.5746+\delta^2 \left(-0.1440\, Ric(v,v)+{\cal O}(\delta)\right),\hspace{1.9cm} \;\;\; D=2, \\
1.6250+\delta^2 \left(-0.0612\, Ric(v,v)-0.0122\, R +{\cal O}(\delta)\right),\;\; D=3, \\
1.6524+\delta^2 \left(-0.0469\, Ric(v,v)-0.0067\, R +{\cal O}(\delta)\right),\;\; D=4,
\end{array}\right.
\label{coeffs}
\end{eqnarray}
where $v$ denotes the unit vector at $p$ in the direction of $p'$ and $R$ is the Ricci scalar 
(trace of the Ricci tensor) at $p$.
The various coefficients in (\ref{coeffs}) arise from the numerical evaluation of exact integral expressions, which depend on $D$
but not on the geometry.\footnote{Note that in $D\! =\! 2$, the
Ricci curvature and Ricci scalar coincide up to a factor of 2.}

Let us emphasize that we follow the same logic as standard lattice field theory with regard to regularization, renormalization 
and continuum limit. This implies that we
are not interested in finite structures (like triangulations or other polygonizations of spacetime) per se,
but only in suitable infinite limits associated with potential quantum gravity theories, where most lattice details will become
irrelevant. An important outcome of our analysis of quantum Ricci curvature on various piecewise flat spaces \cite{qrc1}
was that these so-called lattice artefacts appear to be 
confined to a region $\delta\lesssim 5$, with $\delta$ measured in discrete link units.

\section{The curvature of DT quantum gravity in 2D}
\label{sec:dtqg}

Two-dimensional Euclidean quantum gravity can be defined as the scaling limit of the nonperturbative path integral \cite{bookadj}
\begin{equation}
Z(\lambda)=\sum_{T\in {\cal T}}\frac{1}{C_T}\, {\rm e}^{-S[T]}, \;\;\; S[T]=\lambda N(T),
\label{zzz}
\end{equation}
where $\lambda$ denotes the (bare) cosmological constant.
The triangulations $T$ summed over in (\ref{zzz})
are two-dimensional simplicial manifolds of spherical topology, made of equilateral 
triangles, and $C_T$ denotes the order of the automorphism group of $T$. 
The Einstein-Hilbert action $S$ reduces to a cosmological term proportional to the (discrete) volume
of a configuration $T$, namely, the number $N(T)$ of triangles contained in $T$.\footnote{The
number of edges is $N_1(T)\! =\! \frac{3}{2} N(T)$ and the number of vertices $N_0(T)\! =\!\frac{1}{2} N(T)+2$.}
A common choice, which we will also adopt here, is to work with a particular discrete notion of geodesic distance.
It is measured only between pairs $(q,q')$ of
vertices, where $d(q,q')$ equals the number of edges of the shortest path between $q$ and $q'$, and is therefore also integer-valued.
In line with our remarks at the end of Sec.\ \ref{sec:qrc}, the details of these discretization choices (and others made below) should not
matter in the continuum limit.

The analogue of the average sphere distance on triangulations is given by 
\begin{equation}
\bar{d}(S_p^{\delta},S_{p'}^{\delta})=\frac{1}{N_0(S_p^{\delta})}\frac{1}{N_0(S_{p'}^{\delta})}
\sum_{q\in S_p^{\delta}} \sum_{q'\in S_{p'}^{\delta}} d(q,q'),
\label{sdist_d}
\end{equation}   
where $N_0(S^\delta_p)$ denotes the number of vertices at distance $\delta$ from the vertex $p$. Because of the
branching baby-universe structure of typical DT configurations, the $\delta$-``spheres" $S^\delta$ we defined earlier
below eq.\ (\ref{sdist}) will in general be multiply connected, even for small $\delta$.
One way of making expression (\ref{sdist_d}) into a ``classical" observable is by averaging it over all point pairs,
\begin{equation}
\bar{d}_T(\delta)=\frac{1}{n_T(\delta)}\sum_{p\in T}\sum_{p'\in T} \bar{d}(S_p^{\delta},S_{p'}^{\delta})\, {\boldsymbol\delta}_{d(p,p'),\delta},
\label{aver}
\end{equation}
where the Kronecker delta $\boldsymbol\delta$ enforces the distance $\delta$ between $p$ and $p'$, and
$n_T(\delta)$ denotes the number of point pairs $(p,p')$ in $T$ with $d(p,p')=\delta$. Note that eq.\ (\ref{aver}) includes
an average over directions of the quasi-local quantum Ricci curvature. 

\begin{figure}[th]
\centerline{\scalebox{0.3}{\rotatebox{0}{\includegraphics{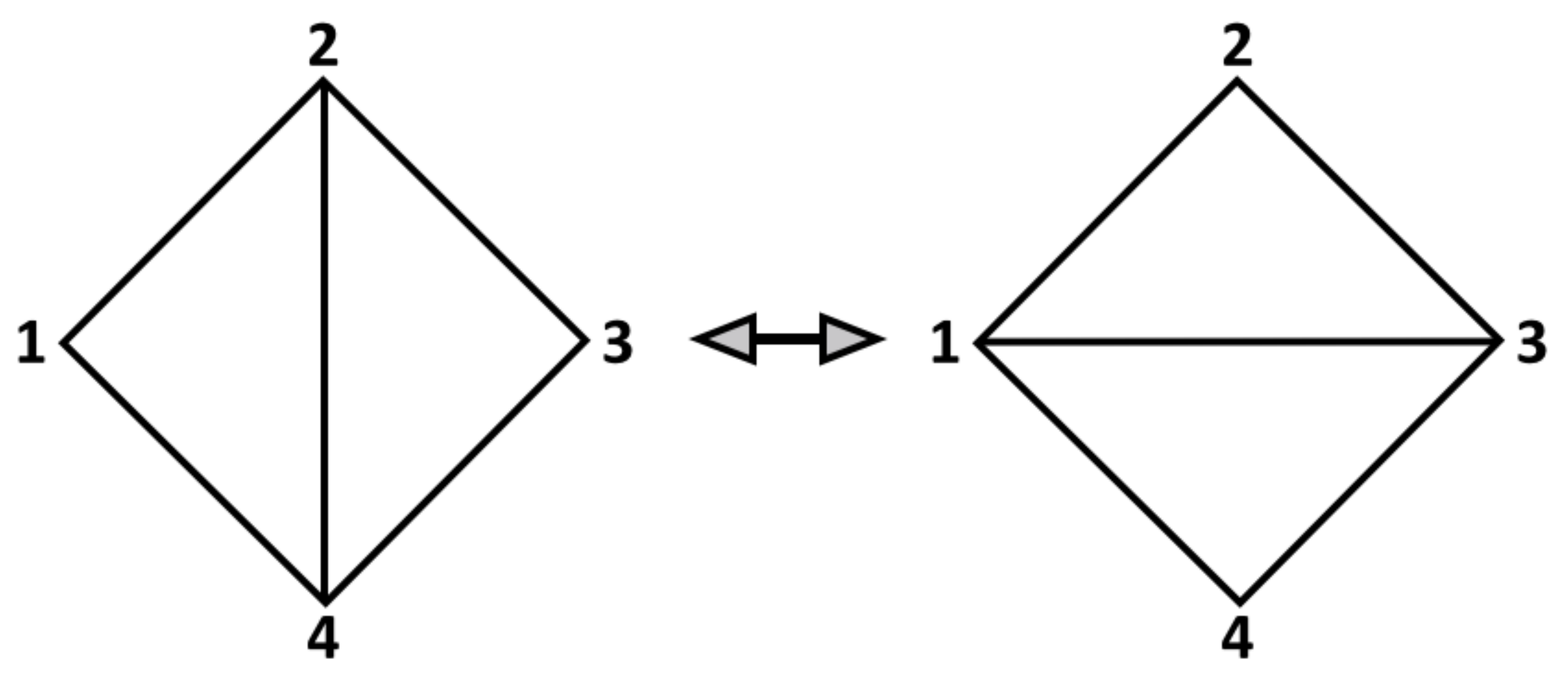}}}}
\caption{The flip move changes the local geometry of a two-dimensional triangulation by substituting a pair of adjacent triangles,
(1,2,4) and (2,3,4), by the triangle pair (1,2,3) und (1,3,4), or vice versa.}
\label{fig:flip}
\end{figure} 

The quantity we have studied with the help of Monte Carlo simulations is the (normalized) expectation value 
$\bar{d}(\delta)/\delta\! :=\!\langle\bar{d}_T(\delta)\rangle_{N}/\delta$ 
in the DT ensemble of (\ref{zzz}), for fixed volume $N$, and specifically its behaviour as $N\rightarrow\infty$.
We used a Metropolis algorithm, with updates by so-called flip moves, consisting in flipping the inner edge of a pair
of adjacent triangles, see Fig.\ \ref{fig:flip}. This move is ergodic in the set of triangulations of fixed topology and fixed volume (see, for
example, \cite{bookadj}). It is always accepted when the algorithm proposes it, provided the resulting triangulation
satisfies the local simplicial manifold conditions. 

\begin{figure}[t]
\centerline{\scalebox{0.5}{\rotatebox{0}{\includegraphics{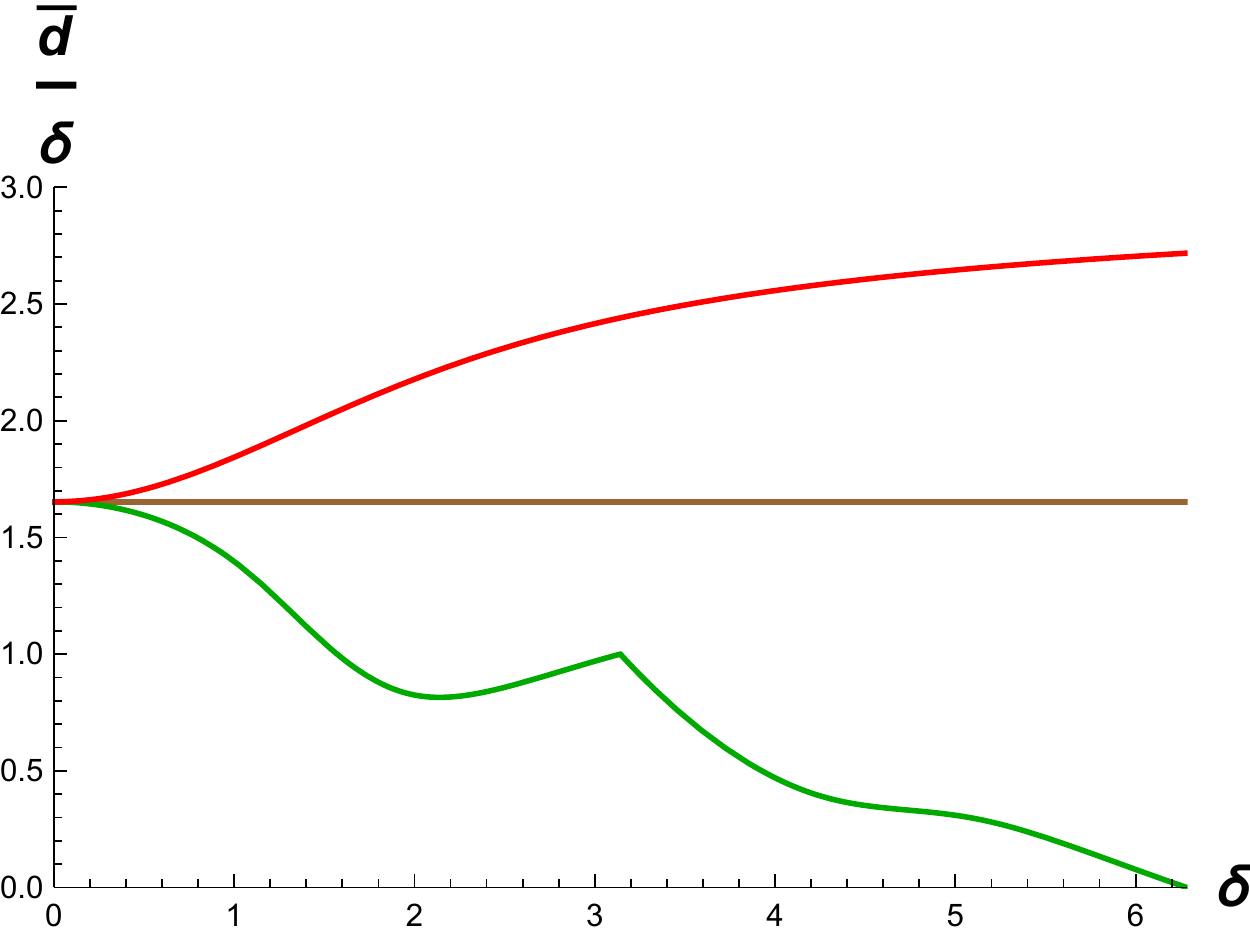}}}}
\caption{The normalized average sphere distance $\bar{d}/\delta$, as function of the scale $\delta\!\in\![0,2\pi ]$, for three constantly
curved manifolds in four dimensions: hyperbolic, with $K_q\! <\! 0$ (top); flat, with $K_q\! =\! 0$ (middle); spherical, with $K_q\! >\! 0$ (bottom). 
The curvature radius $\rho$ of the non-flat spaces has been set to 1. The curves in other dimensions $D$ are 
qualitatively very similar \cite{qrc1}.}
\label{fig:nasd4D}
\end{figure} 
For fixed volume $N$, we performed a double-sampling of $\bar{d}/\delta$ over geometries and point pairs, where a single 
measurement consists of
the following steps: (i) pick a random vertex $p\!\in\! T$ in the triangulation $T$ that has been generated at this stage of
the Metropolis algorithm; (ii) pick a vertex $p'$ randomly from the
sphere of radius $\delta\! =\! 1$ around $p$ and compute $\bar{d}(S_p^{\delta},S_{p'}^{\delta})/\delta$; (iii) repeat step (ii) 
for the same $p$ and a randomly chosen point $p'$ from the $\delta$-sphere around $p$, for $\delta\! =\! 2,3,\dots,15$, yielding
a total of 15 data points for $T$. -- Sweeps between measurements consisted of $4\times 10^7$ suggested updates each. 
We performed 100.000 measurements at volume $N\! =\! 20k$, 80.000 at $N\! =\! 30k$, and 
20.000 each at $N\! =\! 40,60,80,120,160$  and $240k$.

To start with, we wanted to understand whether the measured  
normalized average sphere distance $\bar{d}(\delta)/\delta$ shows any qualitative resemblance
with the behaviour of a smooth manifold of constant (sectional) curvature, at least for some range of $\delta$. 
Spheres, hyperbolic and flat spaces serve as convenient references because of the distinct behaviour of
their quantum Ricci curvature $K_q$, extracted from $\bar{d}(\delta)/\delta\! =\! c_q(1\! -\! K_q(\delta))$, as illustrated
by Fig.\ \ref{fig:nasd4D}. Note that our a priori choice of spherical topology for the triangulations $T$ does not
impose any obvious topological constraints on $K_q(\delta)$, since the quantum Ricci curvature
does not satisfy a Gauss-Bonnet theorem for any value of $\delta$. At any rate, it is unclear what the physical
status of such a theorem would be in a theory where we already know that in the scaling limit the dynamically generated
quantum geometry bears little resemblance to a classical two-dimensional Riemannian manifold.

\begin{figure}[t]
\centerline{\scalebox{0.3}{\rotatebox{0}{\includegraphics{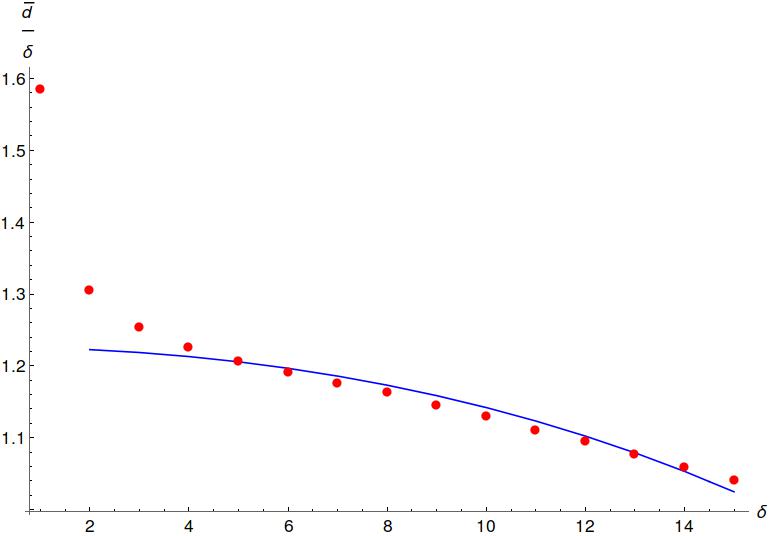}}}}
\caption[Caption for LOF]{Expectation value of the normalized average sphere distance $\bar{d}/\delta$ in dynamical triangulations 
at volume $N\! =\! 40k$, as function of the link distance $\delta$, and best fit of the data for $\delta\! >\! 5$ to a continuum sphere of 
dimension $D\! =\! 4$\,\footnotemark\, (continuous curve) 
and curvature radius $\rho\! =\! 16.7$. (Error bars are smaller than dot size.)}
\label{fig:dbar20k}
\end{figure}

Fig.\ \ref{fig:dbar20k} shows the measured
expectation values of the normalized average sphere distance $\bar{d}/\delta$, taken at volume
$N\! =\! 40k$. For small $\delta$, approaching the mini\-mal value $\delta\! =\! 1$, we
observe a rapid increase in $\bar{d}/\delta$. This seems to be the same short-distance effect we found for 
$\delta\! \lesssim\! 5$ for all of the
piecewise flat geometries investigated in \cite{qrc1}, {\it irrespective of their behaviour for larger $\delta$}, and which we
have already identified as a discretization artefact. For $\delta\! \gtrsim\! 5$, we enter a region of gentler, monotonic decrease,
suggestive of a positive quantum Ricci curvature, corresponding to the initial section of the bottom curve in Fig.\ \ref{fig:nasd4D}.

This motivated our next step, a systematic quantitative comparison of the measured data for $\langle\bar{d}/\delta\rangle$ 
with the corresponding curves for continuum spheres of constant curvature.
In view of the fact that the spectral dimension of DT quantum gravity is $d_S\! =\! 2$ and its Hausdorff dimension
$d_H\! =\! 4$, we have performed fits to spheres with a range of dimensions, $D\! =\! 2$, 3, 4  and 5. 
Unlike in the con\-tinuum, where the constant $c_q^{cont}\! :=\! \lim_{\delta\rightarrow 0}\bar{d}/\delta$ of eq.\ (\ref{qric}) is universal, 
as already mentioned in Sec.\ \ref{sec:qrc}, we established in \cite{qrc1} that the same is not true in the realm of
piecewise flat spaces, where we have identified $c_q\!\equiv\! (\bar{d}/\delta)|_{\delta = 5}$. Using the same identification for the 
DT data, a relative shift between continuum and lattice data is needed to account for the different values of $c_q$.  
Following \cite{qrc1}, we require all curves to go through the point at $\delta\! =\! 5$ and use two alternative methods  
to achieve this, a relative additive shift and a relative multiplicative shift between the continuum and DT data for $\bar{d}/\delta$.
\footnotetext{It will become clear presently why we are not interested exclusively in $D\! =\! 2$.
The analogous curves in dimensions 2, 3 and 5 are qualitatively similar.}

After fixing the relative shift for a data set at given volume $N$, we determine the remaining
free parameter, namely, the effective curvature radius $\rho_{\mathrm{eff}}$ of the sphere that best fits the data.\footnote{The 
sectional curvature of round spheres in any dimension is $1/\rho^2$.} 
Note that the continuum functions $\bar{d}/\delta$ follow universal curves like those depicted in Fig.\ \ref{fig:nasd4D} when 
written in terms of rescaled distances $\delta/\rho$, where $\rho$ is the curvature radius of the spherical or hyperbolic space in question.
Since $\bar{d}/\delta$ at a given point must be computed by numerical integration, to save on computational resources we did this
calculation once for each of 100 evenly spaced values $\delta/\rho\in[0,2\pi ]$, and used linear interpolation to compute
$\bar{d}/\delta$ for arbitrary $\delta$ and $\rho$. The integrations over the sphere pairs were done with Mathematica for $D\!\leq\! 4$,
and with the help of a small program implementing Monte Carlo importance sampling for the case $D\! =\! 5$.
We then performed $\chi^2$-fits on the set of data points with $\delta\! >\! 5$ to determine 
the optimal curvature radius $\rho_{\mathrm{eff}}$ minimising $\chi^2$. 
To this end, we sampled $\rho$-values in discrete intervals of 0.01 for $N\! =\! 20k$ and $30k$ and 0.1 for all larger volumes. 

\begin{figure}[t]
\centerline{\scalebox{0.6}{\rotatebox{0}{\includegraphics{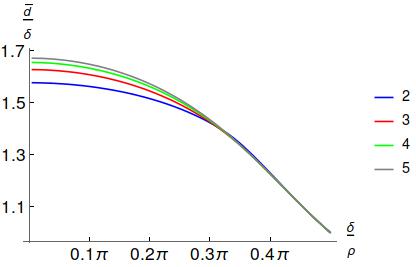}}}}
\caption{Comparing normalized average sphere distances for continuum spheres in dimensions $D\! =\! 2,\dots, 5$ (bottom to top),
covering the range $\delta/\rho \lesssim 0.4\pi$ in normalized distances, which is also explored in the Monte Carlo simulations.}
\label{fig:dimensions}
\end{figure} 
Fig.\ \ref{fig:dbar20k} includes a fit of the data for $N\! =\! 40k$ to a four-dimensional sphere, obtained by using
an additive shift and $\rho_{\mathrm{eff}}\! =\! 16.7$ in units of edge length. 
Given the highly nonclassical nature of the underlying geometry, a perfect fit to the given functional form could have hardly been 
expected, but considering this circumstance the closeness of the fit is still quite remarkable. 
\begin{table}[b!]
\begin{center}
\begingroup
\setlength{\tabcolsep}{0.6\tabcolsep}
\footnotesize
\begin{tabular}{|c c c c c c c c c|}
\hline
\! $N$\! & $D\! =\! 2$, + & $D\! =\! 2$, $\times$ & $D\! =\! 3$, + & $D\! =\! 3$, $\times$ & $D\! =\! 4$, + & $D\! =\! 4$, $\times$ & $D\! =\! 5$, + & $D\! =\! 5$, $\times$ \\
\hline
$20k$ &  12.94(2) &  11.805(15) &  14.04(2) &  12.49(2) &  14.68(3) &  12.86(2) &15.02(3) & 13.07(2)\\
$30k$ &  13.67(2) &  12.52(2) &  15.04(3) &  13.37(2) &  15.76(4) &  13.82(3) &16.16(4) & 14.06(3)\\
$40k$ &  14.28(6) &  13.09(5) &  15.93(8) &  14.08(6) &  16.71(9) &  14.61(7) & 17.14(8) & 14.89(6)\\
$60k$ &  15.15(9) &  13.81(6) &  17.01(12) &  15.07(9) &  17.88(13) &  15.67(10) & 18.38(13) & 15.99(10)\\
$80k$ &  15.97(10) &  14.51(8) &  18.09(12) &  15.97(10) &  19.03(13) &  16.63(10) & 19.56(14) & 16.98(11) \\
$120k$ &  17.18(16) &  15.50(13) &  19.6(12) &  17.21(16) &  20.6(2) &  17.95(17) & 21.2(2) & 18.39(18) \\
$160k$ &  18.6(2) &  16.63(18) &  21.2(3) &  18.6(2) &  22.4(3) &  19.5(2) & 23.1(3) & 19.9(2) \\
$240k$ &  20.0(2) &  17.92(18) &  23.0(3) &  20.1(2) &  24.3(3) &  21.0(2) & 25.0(3) & 21.6(2) \\
\hline
\end{tabular}
\endgroup
\end{center}
\caption{Effective curvature radii $\rho_{\mathrm{eff}}$ obtained from fitting Monte Carlo data to smooth spheres
in various dimensions $D$, using both additive (``+") and multiplicative (``$\times$") shifts, and system sizes of up to $N\! =\! 240k$.}
\label{table1}
\end{table} 

Our results for the effective curvature radii for volumes of up to $240k$, using both additive and multiplicative shifts, and 
fitting to spheres of dimension up to five, are collected in Table \ref{table1}.
Since the shapes of the continuum curves for $\bar{d}/\delta$ in the region sampled are rather similar for different dimensions 
$D$ (c.f.\ Fig.\ \ref{fig:dimensions}), the quality of the
fits, measured in terms of the mean squared deviation, does not depend strongly on $D$, although it does become slightly better as $D$ increases.
Also, using an additive instead of a multiplicative shift improves the fit quality slightly. As an illustration, the mean squared deviation
for $N\! =\! 40k$ and $D\! =\! 5$ is 0.0392 for the multiplicative shift and 0.0379 for the additive shift. The difference is small, but
systematic and of similar magnitude for other volumes and dimensions.

Two more observations can be extracted from Table \ref{table1}. Firstly, the effective curvature radius $\rho_{\mathrm{eff}}$ becomes larger 
when we increase the dimension of the sphere that is being fitted to, while keeping the discrete volume fixed. 
The likely explanation is that the continuum curves for higher $D$ are initially steeper (c.f. Fig.\ \ref{fig:dimensions}), 
which is also achieved by having a larger $\rho_{\mathrm{eff}}$.
Secondly, $\rho_{\mathrm{eff}}$ is systematically larger (on the order of 10\,\!\! --15\%) when using an additive rather than 
a multiplicative shift. At this stage, this is simply a piece of information to be taken on board, because
we do not have good theoretical arguments to prefer one type of shift to the other.

\begin{figure}[t]
\centerline{\scalebox{0.6}{\rotatebox{0}{\includegraphics{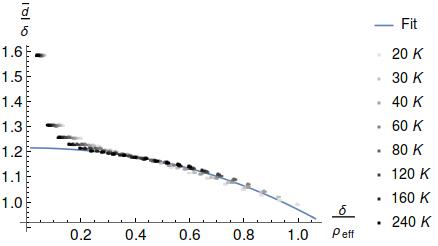}}}}
\caption{Measurements of the normalized average sphere distance as function of the rescaled distance $\delta/\rho_{\mathrm{eff}}$,
for volumes in the range $N\!\in\! [20k,240k]$, with best fit to a five-dimensional continuum sphere. (Error bars too small to be
visible.)}
\label{fig:scale1}
\end{figure} 

Next, we need to investigate the behaviour of the quantum Ricci curvature as a function of the volume $N$, and gather evidence of
whether a scaling limit exists as $N\!\rightarrow\!\infty$. For the comparison with continuum spheres to be meaningful, we expect 
that a rescaling of $\delta$ by the effective curvature radius for given $N$ will be needed.
Working with $D\! =\! 5$ and an additive shift for definiteness, the general situation is illustrated well by
Fig.\ \ref{fig:scale1}, which shows the expectation values
$\langle\bar{d}/\delta\rangle$ as a function of the rescaled distance $\delta/\rho_{\mathrm{eff}}$ for a wide range of discrete volumes.
The optimal shift of the continuum curve was determined by removing all data points with $\delta\! <\! 5$ and then performing 
a joint fit to the remaining data.
Beyond the region of lattice artefacts for small distances, the data points settle down to a common curve in an intermediate regime. 
Their observed slight spread for larger values of $\delta/\rho_{\mathrm{eff}}$ is expected due to the presence of
finite-size effects. This should be most pronounced for the simulations at smaller volume, which is in agreement with our data. 
We conclude that our measurements provide good evidence that in the continuum limit of 2D dynamical triangulations of spherical topology, 
the quantum Ricci curvature $\langle K_q\rangle$ is a well-defined finite quantity with a nontrivial scale dependence, and is positive
in the range of $\delta/\rho$ considered.  

Assuming that the data for the largest volume
$N\! =\! 240k$ represent a good approximation of the infinite-$N$ behaviour of the observable $\langle\bar{d}/\delta\rangle$, the match with the
continuum curve is remarkably good. It is not perfect, due to a slight `overshoot' of the measurements at the largest values of 
$\delta/\rho_{\mathrm{eff}}$ we considered. However, since there is no reason to believe that the DT quantum geometries 
behave like smooth round spheres in terms of {\it all} their metric properties, this is not particularly surprising.

\begin{figure}[t]
\centerline{\scalebox{0.45}{\rotatebox{0}{\includegraphics{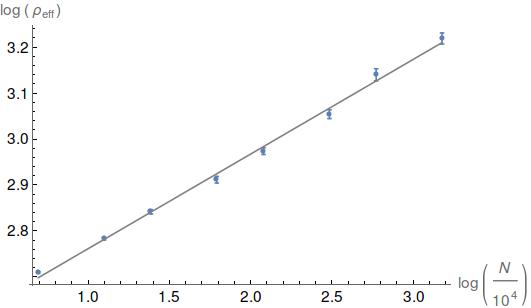}}}}
\caption{Log-log plot of the effective radius $\rho_{\mathrm{eff}}$ as a function of the volume $N$, together with a linear fit,
leading to ${\cal D}\! =\! 4.85(16)$. 
In the case shown, $\rho_{\mathrm{eff}}$ was extracted from fitting to a five-sphere (i.e. $D\! =\! 5$) and using an additive shift.}
\label{fig:fitreff}
\end{figure} 
Since the quality of the sphere fits and the scaling behaviour as a function of volume show only a slight dependence 
on the dimension $D$ of the spheres, we have performed an additional measurement to determine which value of $D$
describes the curvature properties of the quantum geometry best.
The criterion we use is motivated by the property of round $\cal D$-dimensional spheres in the continuum, whose volume $V$
scales with the $\cal D$th power of the curvature radius $\rho$, $V(\rho)\!\propto\! \rho^{\cal D}$.
If the dynamically triangulated quantum geometry indeed resembles a $\cal D$-dimensional sphere globally, one
would expect that its effective curvature radius scales with the corresponding fractional power, that is,
\begin{equation}
\rho_{\mathrm{eff}}\propto N^{\frac{1}{\cal D}}.
\label{fracpower}
\end{equation} 
We have examined all eight cases listed in Table \ref{table1} in search of power law behaviour of the form (\ref{fracpower}).
We look for consistent pairs of $D$ and $\cal D$, where the two dimensions agree.
To extract the dimension $\cal D$, we fitted the data in each case to a straight line in a log-log plot. An example is
depicted in Fig.\ \ref{fig:fitreff}, for $D\! =\! 5$ and using an additive shift in the sphere matching. As illustrated by the case
shown, the quality of the linear fits is generally good. The complete set of results is listed in Table \ref{table2}.
From its entries, we conclude that the only consistent case is that of five dimensions, where $D$ and $\cal D$ 
agree within error bars, for either choice of shift.
\begin{table}[h!]
\begin{center}
\begin{tabular}{|c c c|}
\hline
$D$ from sphere fit & $\cal D$, + & $\cal D$, $\times$\\
\hline
2 & 5.7(3) & 6.0(3) \\
3 & 5.02(17) & 5.21(17) \\
4 & 4.92(17) & 5.04(15) \\
5 & 4.85(16) & 4.94(14) \\
\hline
\end{tabular}
\end{center}
\caption{Dimension $\cal D$ extracted from the scaling law (\ref{fracpower}) for the effective curvature radius
$\rho_{\mathrm{eff}}$, with $\rho_{\mathrm{eff}}$ obtained from fitting to a $D$-dimensional continuum sphere,
using an additive (``+") or multiplicative (``$\times$") shift.}
\label{table2}
\end{table}

Lastly, one interesting aspect of the (quasi-local) quantum Ricci curvature -- inherited from its classical, tensorial counterpart -- is that it allows 
us in principle to capture some directional properties of curvature. Note that this is already true in dimension two. 
We have performed a small exploratory study for $N\! =\! 40k$ and $\delta\! =\! 15$ on two-dimensional dynamical triangulations, 
to try and understand whether the
high degree of anisotropy of the geometry {\it at a given vertex of a given DT configuration} (due to its `baby universe structure')
leaves an imprint on the curvature measurements. To this end, we considered the $\delta$-sphere $S^\delta_p$ around some fixed
vertex $p$ and measured the distribution of the average sphere distance $\bar{d}(\delta)$ to all spheres $S^\delta_{p'}$ with
centre $p'$ on the sphere $S^\delta_p$. In this manner one explores all directions around the point $p$. 
A signal of anisotropic behaviour would be if the average sphere distances clustered around two or more distinct values.
However, after taking suitable averages of the measured directional distributions to get rid of the vertex dependence, we found that the
resulting average distribution is approximately Gaussian with a single peak. It suggests that, at least for distances of the order of 
$\delta/\rho\! \approx\! 1$,
this quantity behaves perfectly isotropically, further illustrating the robustness of quantum Ricci curvature as a quantum observable.

\section{Summary and outlook}
\label{sec:concl}

We set out to evaluate a recently introduced geometric quantum observable, the quantum Ricci curvature, in two-dimensional
quantum gravity defined by dynamical triangulations. The quantity measured directly in the Monte Carlo simulations was
the expectation value $\langle\bar{d}/\delta\rangle$ of the normalized average sphere distance, from which the quantum
Ricci curvature $\langle K_q(\delta)\rangle$ can be extracted via a quantum analogue of eq.\ (\ref{qric}). 

A qualitative inspection of the data $\langle\bar{d}/\delta\rangle (\delta)$ signalled a positive curvature in the
$\delta$-range considered. This motivated a further quantitative comparison with smooth classical model spaces
of constant positive curvature, given by round spheres of dimension $D$.  For given $D\!\in\!\{ 2,3,4,5\}$, we extracted the curvature radius of
the sphere that best fitted the Monte Carlo data, for a total of eight different system sizes $N\leq 240.000$.
The effective curvature radii thus obtained (Table \ref{table1}) allowed us to perform a scaling analysis for given $D$, 
by plotting the measured data as function of the rescaled distances $\delta/\rho_{\mathrm{eff}}$. 
The rescaled data display only a mild $N$-dependence, and provide good evidence for the existence of a universal
curve $\langle\bar{d}/\delta\rangle$ as $N\!\rightarrow\!\infty$. We noted that the quality of the fits to curves for continuum spheres 
becomes slightly better as the dimension $D$ increases. Unlike what one might have expected na\"ively in a theory of
two-dimensional quantum gravity, $D\! =\! 2$ therefore appeared to be the least-preferred of the integer values we considered. 

To extract a `best dimension' associated with the quantum Ricci curvature construction, 
we then investigated a complementary criterion for measuring a dimension, by looking at the dependence of 
the effective curvature radii on the system size.   
Since the Hausdorff dimension of DT quantum gravity is four, a natural conjecture would be that
the preferred sphere dimension to emerge from the curvature analysis is also four. Instead, it turned out that the preferred,
self-consistent dimension associated with the numerical data gathered is five.
This noncanonical behaviour underscores the fact that the dynamically generated quantum geometry of the
model is highly nonclassical and does not resemble any given smooth geometry. 

We cannot offer a more specific interpretation of the value five at this stage, 
because DT quantum gravity in two dimensions is currently the only nonperturbative quantum model whose 
quantum Ricci curvature has been analyzed. In addition, the model
does not have a conventional classical limit one could use for comparison, because there is no classical theory of general 
relativity in two spacetime dimensions.
It would be very interesting to understand whether the quantum Ricci curvature can be derived analytically in one of the various
descriptions of two-dimensional Euclidean quantum gravity that are available \cite{2dgrav,bookadj}. 

An important aspect of the present work is the demonstration that the quantum Ricci curvature can be implemented 
in nonperturbative quantum gravity in a rather straightforward way. Although we did not consider the full-fledged theory
in four dimensions, the two-dimensional model we examined is of particular interest because of the highly 
nonclassical nature of its quantum geometry, far removed from the spaces we considered earlier \cite{qrc1}.
As an added `bonus' we found the unexpected result that its quantum Ricci curvature follows 
closely that of a smooth five-sphere.

In summary, we have established by way of a nontrivial example that {\it quantum Ricci curvature} is a 
well-defined quantum observable, giving us a new, independent way to quantify the 
properties of quantum geometry.   
Given the general dearth of observables in quantum gravity, this is an important result. 
It opens a variety of avenues for further exploration, most importantly,
the application to full nonperturbative (CDT) quantum gravity in four spacetime dimensions, as outlined in Sec.\ \ref{sec:need}, which is
currently under way. However, any formulation of quantum gravity that uses discrete building blocks 
with suitable metric properties to describe space(time) should provide a natural habitat for quantum Ricci curvature. 
Among the other interesting aspects to be understood, also in lower dimensions, is the interplay between local and global, more specifically,
the influence of global topology and boundary conditions on the quantum Ricci curvature. We hope to return to this issue in the near future.

\subsection*{Acknowledgments} 
This work was partly supported by the research program
``Quantum gravity and the search for quantum spacetime" of the Foundation for Fundamental Research 
on Matter (FOM, now defunct), financially supported by the Netherlands Organisation for Scientific Research (NWO).

\end{document}